\def\be{\begin{equation}}
\def\ee{\end{equation}}
\def\ll{\label}
\documentstyle[12pt,a4]{article}
{\LARGE \bf\author{{\sc  Ladislav   Hlavat\'{y}}
\thanks{Postal   address:
B\v{r}ehov\'{a} 7, 115 19
Prague 1, Czech Republic. E-mail: hlavaty@br.fjfi.cvut.cz}
\\ {\it Department  of  Physics,}
\\ {\it  Faculty  of  Nuclear  Sciences  and
Physical Engineering}}
\title{Open spin chains and quadratic algebras}
\date{15 December 1993}

\begin{document}

\maketitle
hep-th/9312130 \hskip 8cm PRA-HEP-93/33
\vskip 2cm

\abstract{Algebraic framework for construction of a  commuting set
of operators that can be interpreted as
integrals of motion of the open spin chain with boundary
conditions and nearest neighbour interaction is investigated.}
\newpage

In  the  end  of  the  seventies  quantum inverse scattering
method was developed (for a review see
e.g. \cite{takhfad}). One of the
systems where it was applied was {\em periodic} spin chain with the
nearest neighbour interaction. The
algebra, from which the hamiltonian of this system as well as its
integrals of motion were derived, is defined by the relations
\be R_{12}(u_1-u_2)L_1(u_1)L_2(u_2)=L_2(u_2)L_1(u_1)R_{12}(u_1-u_2)
\ll{rll} \ee
where $R$ is a matrix function $R:U\rightarrow End(V_0\otimes V_0)$
satisfying the Yang--Baxter equation
(YBE)
\be R_{23}(u_2-u_3)R_{13}(u_1-u_3)R_{12}(u_1-u_2)=
R_{12}(u_1-u_2)R_{13}(u_1-u_3)R_{23}(u_2-u_3)\ll{ybe} \ee
and the range $U$ of the "spectral  parameters" $u$ is usually the
field of complex numbers
${\bf C}$.

In 1988, Sklyanin proposed a method for constructing solvable
models of quantum {\em open} (i.e. non--periodic)
 {\em spin chains} \cite{sklopsp}. The method is based on
reflection-type algebras given by the relations
\[ R_{12}(u_1-u_2)M_1(u_1)R_{12}(u_1+u_2)M_2(u_2)
=M_2(u_2)R_{12}(u_1+u_2)M_1(u_1)R_{12}(u_1-u_2)\]
\[ R_{12}(u_2-u_1)K_1^{t_1}(u_1)R_{12}(-u_1-u_2-2\eta )
 K_2^{t_2}(u_2)              = \hskip 4cm \]
\[ \hskip 4cm
 K_2^{t_2}(u_2)R_{12}(-u_1-u_2-2\eta )
 K_1^{t_1}(u_1)R_{12}(u_2-u_1)\]
The matrix $R$, beside the YBE, satisfies conditions
\[ P_{12}R_{12}(u)P_{12}=R_{12}(u),\]
\[ R_{12}^{t_1}(u)=R_{12}^{t_2}(u),\]
\[ R_{12}(u)R_{12}(-u)=\rho (u){\bf 1} _{12},\]
\[ R_{12}^{t_1}(u)R_{12}^{t_1}(-u-2\eta )=\tilde \rho
(u){\bf 1}_{12},\]
 where $t_1,t_2$ mean the transposition in the first,
respectively second pair of indices.
Some of these conditions were later weakened \cite{meznep} but stil
they remained rather restrictive.
The purpose of the present paper is to present a more general
construction of open spin chains.

 The starting point is
associative algebra
generated by elements $M_i^j(u),K_i^j(u)$,
 $i,j \in \{1,\ldots ,d_0=dimV_0\}$ satisfying quadratic relations
\be A_{12}(u_1,u_2)M_1(u_1)B_{12}(u_1,u_2)M_2(u_2)=
M_2(u_2)C_{12}(u_1,u_2)M_1(u_1)D_{12}(u_1,u_2),
\ll{ambm} \ee
\be \tilde A_{12}(u_1,u_2)K_1^{t_1}(u_1)\tilde B_{12}(u_1,u_2)
K_2^{t_2}(u_2)=K_2^{t_2}(u_2)
\tilde C_{12}(u_1,u_2)K_1^{t_1}(u_1)\tilde D_{12}(u_1,u_2).
\ll{akbk} \ee
where $A, B,\ldots,\tilde D$ are matrix functions $U\times
U\rightarrow End(V_0\times V_0)$ i.e.
$A_{12}(u_1,u_2)=A_{i_1i_2}^{j_1j_2}(u_1,u_2)$,
$i_1,i_2,j_1,j_2 \in \{1,\ldots ,d_0=dimV_0\}$
and similarly for $B, C,\ldots,\tilde D$.
Algebras of this type were investigated in another
context in \cite{frimai}.

Our first task is to identify the conditions on the numerical matrices
$A,B,\ldots ,\tilde D$
that guarantee the existence of a commuting subalgebra that, when
represented, will provide us with a commuting set of operators
that eventually can be interpreted as the integrals of motion of
a quantum system.

{\em Theorem 1:} Let $\cal A$ is the associative algebra generated by
elements $M_i^j(u)$, $K_i^j(u)$, relations
(\ref{ambm}), (\ref{akbk}) and
\be M_1(u_1)K_2(u_2) = K_2(u_2)M(u_1).\ll{mkkm} \ee
 If  the matrices $\tilde  A, \tilde
B, \tilde C, \tilde D, $ are related
to $A, B, C, D$ by

\be  \tilde A_{12}(u_1,u_2)=(A_{12}^{t_1t_2}(u_1,u_2))^{-1} ,
\ \  \tilde D_{12}(u_1,u_2)=(D_{12}^{t_1t_2}(u_1,u_2))^{-1} ,
\ll{atdt} \ee
\be \tilde B_{12}(u_1,u_2)={({(B_{12}^{t_1}(u_1,u_2))}^{-1})}^{t_2} ,
\ \  \tilde
C_{12}(u_1,u_2)={({(C_{12}^{t_2}(u_1,u_2))}^{-1})}^{t_1},
\ll{btct} \ee
then the elements $t(u)=K_i^j(u)M_j^i(u)=tr[K(u)M(u)]$
form a commutative
subalgebra
 i.e.
\[ [t(u_1),t(u_2)] = 0. \]

{\em Remark:} Note that there are no restrictions on $A,B,C,D$.

{\em Proof:} Repeats the steps of \cite{sklopsp}. Denoting
$K_1\equiv K_1(u_1)$ , $K_2\equiv K_2(u_2)$ ,
$M_1\equiv M_1(u_1)$ , $M_2\equiv M_2(u_2)$ ,
$A_{12}\equiv A_{12}(u_1,u_2), \ldots ,
 D_{12}\equiv D_{12}(u_1,u_2)$
and using the properties of the trace
\[tr X^tY^t = tr XY,\ \ tr(XY^t)=tr(X^tY) \]
 and relations (\ref{ambm})--
(\ref{btct}) we get
\[ t(u_1)t(u_2) =tr_1(K_1M_1)tr_2(K_2M_2)
=\ldots  =tr_{12}(K_1^{t_1}K_2M_1^{t_1}M_2) \]
\[ = tr_{12}(K_1^{t_1}K_2\tilde B_{12}^{t_2}B_{12}^{t_1}
  M_1^{t_1}M_2) =tr_{12}[{(K_1^{t_1}\tilde B_{12}K_2^{t_2})}^{t_2}
  {(M_1B_{12}M_2)}^{t_1}]=  \]
\[  tr_{12}[{(K_1^{t_1}\tilde B_{12}K_2^{t_2})}^{t_1t_2}
\tilde A_{12}^{t_1t_2}A_{12}
  {(M_1B_{12}M_2)}]
 = tr_{12}[{(\tilde A_{12}K_1^{t_1}\tilde B_{12}K_2^{t_2})}^{t_1t_2}
(A_{12}  {M_1B_{12}M_2)}]  \]
\[=  tr_{12}[{(K_2^{t_2}\tilde C_{12}K_1^{t_1}\tilde D_{12})}
  {(M_2C_{12}M_1D_{12})}^{t_1t_2}]
 =tr_{12}[{(K_2^{t_2}\tilde
C_{12}K_1^{t_1})}^{t_1}{(M_2C_{12}M_1)}^{t_2}]
\] \[ =tr_{12}[K_2^{t_2}K_1M_2^{t_2}M_1]
 = tr_{12}[K_2^{t_2}M_2^{t_2}K_1M_1]         =t(u_2)t(u_1)
\]
Q.E.D

The fundamental property that enabled to construct the operators
describing integrals of motion of  periodic spin chains was
the possibility to define a coproduct in the algebra
(\ref{rll}) because the commuting operators then could be
expressed in the form
\[ t(u) = tr[L_{(N)}(u)L_{(N-1)}(u)\ldots L_{(1)}(u)] \]
where $L_{(j)}$ were matrices of operators
acting nontrivially only in the space
of the $j$--th spin. However, it seems that it is not posible (in the
unbraided categories)  to define a coproduct  in the algebra
$\cal A$. Nevertheless, we can use the
algebra for the construction of spin chain operators due to the
following covariance property.

{\em Theorem  2:}  Let  $\cal  B$  is  the  algebra  generated  by
$L(u)=L_i^j(u), N(u)=N_i^j(u)$,   $i,j \in \{1,\ldots ,d_0=dimV_0\}$
and relations
\be A_{12}(u_1,u_2)L_1(u_1)L_2(u_2) = L_2(u_2)L_1(u_1)A_{12}(u_1,u_2)
\ll{all} \ee
\be   D_{12}(u_1,u_2)N_1(u_1)N_2(u_2)    =
N_2(u_2)N_1(u_1)D_{12}(u_1,u_2)
\ll{dnn} \ee
\be   N_1(u_1)B_{12}(u_1,u_2)L_2(u_2)    =
L_2(u_2)B_{12}(u_1,u_2)N_1(u_1)
\ll{nbl} \ee
\be L_1(u_1)C_{12}(u_1,u_2)N_2(u_2) = N_2(u_2)C_{12}(u_1,u_2)L_1(u_1).
\ll{lcn} \ee
Then

1) the algebra $\cal B$ can be turned into bialgebra by coproduct
\be      \Delta(L_i^j(u))=L_i^k(u)\otimes      L_k^j(u)      ,
\ \ \Delta(N_i^j(u))=N_k^j(u)\otimes N_i^k(u)\ll{cop} \ee
and counit
\be  \epsilon  (L_i^j(u))=  \delta_i^j,  \epsilon (N_i^j(u))=
\delta_i^j. \ll{coun}\ee

2)  the algebra  $\cal M$  generated by  the $M_i^j(u)$  and
relations (\ref{ambm}) is $\cal B$--comodule
algebra. The coaction on $\cal M$ is given by
\be \beta : {\cal M} \rightarrow {\cal M} \otimes {\cal B}, \ \
\beta (M_i^j(u))=M_k^l(u)\otimes             L_i^k(u)N_l^j(u)
\ll{coac}\ee
that with a slight abuse of notation can be written as
$\beta(M)=\tilde M =LMN$.

{\em Remark:} A similar covariance algebra can be defined for the
algebra $\cal K$ generated by $K_i^j(u)$.

{\em Proof:}  It is  straightforward to  check that  the relations
(\ref{all}-\ref{lcn})       are        invariant       under
(\ref{cop},\ref{coun}). Proof of invariance of (\ref{ambm})
under (\ref{coac}) is
\[           A_{12}\tilde M_1B_{12}\tilde M_2           =
A_{12}L_1M_1(N_1B_{12}L_2)M_2N_2                           =
(A_{12}L_1L_2)M_1B_{12}N_1M_2N_2                           =\]
\[L_2L_1(A_{12}M_1B_{12}M_2)N_1N_2                           =
L_2L_1M_2C_{12}M_1(D_{12}N_1N_2) =\]
\[L_2M_2(L_1C_{12}N_2)M_1N_1D_{12}                           =
(L_2M_2N_2)C_{12}(L_1M_1N_1)D_{12} =  \tilde M_2C_{12}\tilde
M_1D_{12}.\] Q.E.D. (We have again deleted  the $(u_1,u_2)-$dependence
in the above formulae.)

Importance of the Theorem 2 is in the fact that it gives a
possibility to
define a set of commuting operators composed from the operators
acting nontrivially only in spaces $V_i$ of single spin states.
Indeed, if $\rho_i$ are representation of $\cal B$ on spaces $V_i$,
$i=1,\ldots,N$, then
\[\hat{L}(u):=(\rho_1\otimes  \rho_2\otimes  \ldots  \otimes
\rho_N)
\circ (\Delta^{N-1})(L(u)), \]

\[\hat{N}(u):=(\rho_1\otimes \rho_2\otimes \ldots \otimes \rho_N)
\circ (\Delta^{N-1})(N(u)) \]
are operators that represent the algebra $\cal B$ on
${\cal H} \equiv V_1\otimes V_2\otimes\ldots\otimes V_N$
that is the Hilbert space of the system of N
spins and the operators $\hat{L}(u),\hat{N}(u)$ can be written as
\be \hat L(u) =\hat{L}_{(N)}(u)\hat{L}_{(N-1)}(u)\ldots
\hat{L}_{(1)}(u)\ll{repl} \ee
\be \hat N(u) =\hat{N}_{(1)}(u)\hat{N}_{(2)}(u)\ldots
\hat{N}_{(N)}(u)\ll{repn} \ee
where
\[\hat{L}_{(j)}(u)={\bf 1}\otimes \ldots \otimes {\bf
1}\otimes\rho_j(L(u))
\otimes {\bf 1}\otimes\ldots\otimes {\bf 1} \]
\[\hat{N}_{(j)}(u)={\bf 1}\otimes \ldots \otimes {\bf
1}\otimes\rho_j(N(u))
\otimes {\bf 1}\otimes\ldots\otimes {\bf 1} \].

The representations of $\cal B$ on $V_i$ such that $dimV_i=dimV_0$
follow from

{\em Theorem 3:} Let there are
$\alpha, \delta \in U$ such that the Matrices $A,B,C,D$
satisfy the equations
\be A_{12}(u_1,u_2)A_{13}(u_1,\alpha)A_{23}(u_2,\alpha)
 =A_{23}(u_2,\alpha) A_{13}(u_1,\alpha)A_{12}(u_1,u_2)
\ll{aaa} \ee
\be D_{12}(u_1,u_2)D_{13}(u_1,\delta)D_{23}(u_2,\delta)
 =D_{23}(u_2,\delta)D_{13}(u_1,\delta)D_{12}(u_1,u_2)
\ll{ddd} \ee
\be D_{13}(u_1,\delta)B_{12}(u_1,u_2)A_{23}(u_2,\alpha)
 =A_{23}(u_2,\alpha)B_{12}(u_1,u_2)D_{13}(u_1,\delta)
\ll{dba} \ee
\be A_{13}(u_1,\alpha)C_{12}(u_1,u_2)D_{23}(u_2,\delta)
 =D_{23}(u_2,\delta)C_{12}(u_1,u_2)A_{13}(u_1,\alpha)
\ll{acd} \ee
for all $u_1,u_2 \in U$.
(note the unusual order of indices in (\ref{dba},\ref{acd})).

Then  the  map  $\rho_{\alpha  \delta} : End
(V_0)\otimes
{\cal B} \rightarrow End(V_0\otimes V_0)$
\be [\rho_{\alpha \delta}(L_k^j(u))]_m^n=
A_{km}^{jn}(u,\alpha)\ll{repl0} \ee
\be [\rho_{\alpha \delta}(N_k^j(u))]_m^n=
D_{km}^{jn}(u,\delta)\ll{repn0}\ee
is a representation of the algebra $\cal B$ on $V_i$ such that
$dim V_i=dimV_0$.

{\em      Proof:}     Direct      check     of     relations
(\ref{all})--(\ref{lcn})         by         means         of
(\ref{aaa})--(\ref{ddd}).

{\em Remark :} Note that full Yang--Baxter type equations are
not
required in the theorem. It is suficient if they are satisfied
for single $(\alpha,\delta)\in U\times U$.

 If we find a representation $\sigma$ of $\cal A$ on
${\cal H}$,
then due to the Theorems 1,2 we get the set of commuting operators
on ${\cal H}$
\[\hat{t} (u)=Tr[\sigma(K(u))\hat{L}(u)\sigma(M(u))\hat{N}(u)] \]
Assuming that there are numerical matrices $m(u),
k(u) \in End(V_0)$ that satisfy the relations (\ref{ambm}),
(\ref{akbk}), we can choose
\[ \sigma(M_i^j(u)) = m_i^j(u){\bf 1}_{\cal H}
,\ \sigma(K_i^j(u)) = k_i^j(u){\bf 1}_{\cal H} \]
and then
\be \hat{t}(u) =tr[k(u)\hat{L}_{(N)}(u)\ldots\hat{L}_{(1)}(u)
m(u)\hat{N}_{(1)}(u)\ldots\hat{N}_{(N)}(u)] \ll{rept} \ee
where the operator matrices $L_{(k)}$ and $N_{(k)}$
act nontrivailly only in the
$k$-th factor of the space ${\cal H}
=V_0\otimes V_0\otimes\ldots\otimes V_0$.

The last goal we want to achieve is finding the hamiltonian $H$ of
the open chain system with the nearest neighbour interaction and
boundary terms.

{\em Theorem 4:} Let there is one--dimensional representation of
the  algebra $\cal A$ by
numerical  matrices  $m(u)$,  $k(u)$ and the  representation of
$\cal B$  on $V_i$ is  $\rho_i = \rho_{\alpha,  \delta}$ for
all $i \in \{1,\ldots,N\}$.

If there is $u_0 \in U$ such that
\be A_{12}(u_0,\alpha)=\kappa P_{12},\ \ D_{12}(u_0,\delta)=\lambda
P_{12}, \ \
m(u_0)=\mu {\bf 1} \ll{reg} \ee
where $\kappa,\lambda,\mu$ are constants and $P$ is the
permutation matrix, then
\be \hat t
(u_0)=\mu(\kappa\lambda)^N tr[ k(u_0)] \ll{tuo} \ee
 and
\be \frac{d\hat{t}}{du} (u_0) =(\kappa\lambda)^N\mu \{ Htr[k(u_0)]
+tr[\frac{dk}{du}(u_0)]\}   \ll{ham1} \ee
\be H=\sum_{n=1}^{N-1}H_{n,n+1} +
\mu^{-1}\frac{dm_{(1)}}{du}(u_0)+[tr\ k(u_0)]^{-1}tr_0[
k_0(u_0)H_{N,0}]  \ll{ham} \ee
\be H_{n,n+1}=\lambda^{-1}\frac{dD_{n,n+1}}{du}(u_0,\delta)P_{n,n+1}
 +
\kappa^{-1}P_{n,n+1}\frac{dA_{n,n+1}}{du}(u_0,\alpha)  \ll{hamm} \ee

{\em Proof:} From (\ref{repl}), (\ref{repn}) and
(\ref{repl0})--(\ref{reg}) we get
\be \hat{L}(u)=\kappa^NP_{0N}P_{0n-1},\ldots,P_{01},\ \
\hat{N}(u)=\lambda^NP_{01}P_{02},\ldots,P_{0N}\ll{lu0} \ee
wherefrom (\ref{tuo}) immediately follows. Similarly,
(\ref{ham1})--(\ref{hamm}) is obtained by differentiating
(\ref{rept}) and using (\ref{lu0}) and the identity on
$End(V_0^{\otimes
N+1})$
\[ P_{0,n+1}X_{0n}=X_{n+1,n}P_{0,n+1}. \]

In the conclusion, we have shown that the {\em open spin chains
with the nearest neighbour interaction
can be constructed from rather general quadratic
algebras} defined by matrix functions  $A,B,C,D$ that

1) satisfy the equations (\ref{aaa},\ref{ddd},\ref{dba},\ref{acd}),

2) admit numerical matrices $m(u), k(u)$ that satisfy
(\ref{ambm},\ref{akbk},\ref{atdt},\ref{btct}).

3) together with $m(u)$ satisfy the regularity condition
(\ref{reg}).


\begin{thebibliography}{99}
\bibitem{takhfad} Takhtajan L.A. and L.D. Faddeev, Uspekhi Mat.
Nauk 34 (1979) 13 {\em in Russian}, Russian Math. Survey 34 (1979)
11 {\em English transl}.
\bibitem{sklopsp}  Sklyanin E, J. Phys. A 21 (1988) 2375
\bibitem{meznep}  Mezincescu L and R I  Nepomechie, J. Phys. A 24
(1991) L17
\bibitem{frimai} Friedel L and J M Maillet, Phys. Lett. B 262
(1991) 278

\end{thebibliography}
\end{document}